\documentclass[11pt]{article}  
\usepackage{amssymb}

  \newcommand{\drm}{{\rm d}}
  
\newcommand{\h}{\hspace*{0.cm}}  
\newcommand{\vs}{\vspace*}  
\newcommand{\noi}{\noindent}
\newcommand{\pa}{\partial}

\newcommand{\nc}{\newcommand}
\nc{\bb}{\begin{equation}} \nc{\ee}{\end{equation}}
\nc{\qq}{\qquad\qquad} \nc{\munu}{{\mu\nu}}
\nc{\dis}{\displaystyle} \nc{\um}{{1\over 2}}
\nc{\R}{{\rm I\!\!R}} \nc{\C}{{\rm I\!\!\!C}}
\nc{\ug}{\; = \;}  \nc{\Hc}{{\cal H}} \nc{\Sc}{{\cal S}}
\nc{\Lc}{{\cal L}} \nc{\LcN}{{\cal L}^{(N)}} \nc{\Lczero}{{\cal L}^{(0)}}
\nc{\Hbf}{\mbox{\boldmath $H$}} \nc{\Ebf}{\mbox{\boldmath $E$}}
\nc{\Bbf}{\mbox{\boldmath $B$}} \nc{\xbf}{\mbox{\boldmath $x$}}
\nc{\lbf}{\mbox{\boldmath $l$}} \nc{\jbf}{\mbox{\boldmath $j$}}
\nc{\abf}{\mbox{\boldmath $a$}} \nc{\sbf}{\mbox{\boldmath $s$}}
\nc{\imp}{\mbox{\boldmath $p$}} \nc{\vbf}{\mbox{\boldmath $v$}}
\nc{\Fbf}{\mbox{\boldmath $F$}} \nc{\rbf}{\mbox{\boldmath $r$}}
\nc{\dox}{\dot{x}} \nc{\ddox}{\ddot{x}}
\nc{\dddox}{{\stackrel{\ldots}{x}}} \nc{\dop}{\dot{p}}
\nc{\dov}{\dot{v}} \nc{\ddov}{\ddot{v}}
\nc{\doa}{\dot{a}} \nc{\ddoa}{\ddot{a}}
\nc{\dddoa}{{\stackrel{\ldots}{a}}} \nc{\ddddoa}{{\stackrel{....}{a}}}
\nc{\rd}{{\rm d}} \nc{\dtau}{{\rd\tau}} \nc{\dt}{{\rd t}}
\nc{\aii}{{a^{(\rm 2i)}}}
\nc{\doabf}{\mbox{\boldmath ${{\stackrel{.}{a}}}$}}
\nc{\ddoabf}{\mbox{\boldmath ${{\stackrel{..}{a}}}$}}
\nc{\dddoabf}{\mbox{\boldmath $\dddoa$}}
\nc{\ddddoabf}{\mbox{\boldmath $\ddddoa$}}
\nc{\omz}{\omega_0} \nc{\kb}{\overline{k}}
\nc{\dosbf}{\mbox{\boldmath ${{\stackrel{.}{s}}}$}}
\nc{\omf}{{\frac{\omega}{\omz}}} \nc{\tz}{\tau_0}

\begin{document}

\title{Deriving spin within a discrete-time theory}

\author{Erasmo Recami \ and \ Giovanni Salesi}

\date{}
\maketitle
\begin{center}
{{\em Universit\`a Statale di Bergamo, Facolt\`a di
Ingegneria,\\
viale Marconi 5, 24044 Dalmine (BG), Italy}\\
and\\
{\em Istituto Nazionale di Fisica Nucleare--Sezione di Milano, Italy\\
via Celoria 16, 20133 Milan, Italy}\\
\ \\
e-mail: {\em recami@mi.infn.it; \ salesi@unibg.it}}
\end{center}

\vs{1cm}

\begin{abstract}
\noindent We prove that the classical theory with a discrete time (chronon)
is a particular case of a more general theory in which
spinning particles are associated with generalized Lagrangians containing
time-derivatives of any order (a theory that has been called
``Non-Newtonian Mechanics"). As a consequence, we get, for instance, a
classical kinematical derivation of Hamiltonian and spin vector for
the mentioned chronon theory (e.g., in Caldirola et al.'s formulation).
Namely, we show that the extension of classical mechanics obtained
by the introduction of an elementary time-interval does actually
entail the arising of an intrinsic angular momentum; so that it may
constitute a possible alternative to string theory in order to account for
the internal degrees of freedom of the microsystems.

\


\end{abstract}

\

\section{Introduction: An elementary quantum of time in particle physics}

The concept of an elementary {\em time}-duration (a ``quantum of time")
has recently returned into fashion in GUTs\cite{GUTs}, in String
Theories\cite{String}, in Quantum Gravity\cite{QG}, and in the approaches
regarding spacetime either as a sort of quantum ether, or as a spacetime
foam\cite{Foam}, or as endowed with a non-commutative geometry (like in
Deformed or Double Special Relativity\cite{NCG}\footnote{In the first of refs.\cite{NCG}
it is stated that ``the special role of the time coordinate in the structure of 
$k$-Minkowski spacetime forces one to introduce an element of discretization in 
the time direction: the time derivative of time-to-the-right-ordered functions is 
indeed standard (just like the $x$-derivative of time-to-the-right-ordered 
functions is standard), but it is a standard $\lambda$-discretized derivative 
(whereas the $x$-derivative of time-to-the-right-ordered functions is a standard 
continuous derivative).''}). In particolar, the M-theory
and the Loop Quantum Gravity (as in its version with semiclassical spin-network
structure of spacetime)\cite{LQG} lead even to a discrete {\em space-time}, in which a
fundamental time-scale (or, equivalently, a mass-energy scale) naturally arises,
besides $\hbar$ and $c$.

\h It is known, and we shall see it again below, that the time-discretization
implies for an elementary object an ``internal'' motion associated with
microscopic space distances: So that one can expect for such an elementary
particle an extended-like, rather than a pointlike, structure.
This is a good since, indeed, even the {\em classical} theory of a pointlike
charged particle leads to obvious divergencies, only seemengly overcome by
the renormalization tecniques (indeed, in QED the infinities do actually persist).
Moreover, a fundamental length is known to be directly linked to the existence
of the energy cut-off needed for avoiding the so-called ultraviolet
catastrophes in any quantum field theories.

\h One of the first, and simplest, theories which assumed a priori a minimum
time interval was Caldirola's theory of the
electron\cite{CALREV4}, based on the existence of an
elementary proper-time duration: the so-called {\em chronon}.

\h Such a finite difference theory possesses rather good characteristics:
for instance, it succeeds  ---already at the classical level--- in forwarding
a solution for the motion of a particle endowed
with a non-negligible charge in an external electromagnetic field, overcoming
all the known difficulties met by Abraham--Lorentz's and Dirac's approaches
(and even allowing a clear answer to the question whether a free falling
charged particle does or does not emit radiation); while ---at the quantum
level,--- it yields a remarkable mass spectrum for leptons.

\section{About the ``chronon" theory}

Let us recall that Caldirola's theory seems, moreover, to explain the
origin of the ``classical (Schwinger's) part'', $e\hbar/2mc\cdot\alpha/2\pi =
e^3/4\pi mc^2$, of the anomalous magnetic momentum of the electron.

\h In the classical version of his theory, however,
Caldirola excluded a priori the existence of spin contributions in
his chronon approach. By contrast, we are going to demonstrate {\em the
rising, even within the classical chronon theory, of an intrinsic angular
momentum}.

\h First of all, let us describe, by starting, e.g., from refs.\cite{CALREV4},
how the chronon theory allows overcoming ---as we were saying---
well-known problems like the so-called ``pre-accelerations" and ``run-away
solutions'' of the Abraham-Lorentz-Dirac equation for the electron, and
furnishing a clear solution to the ambiguities associated with the ``hyperbolic
motion".

\h If $\rho$ is the charge density of a particle on which an external
electromagnetic field acts, the famous Lorentz's force law
\begin{equation}
\Fbf = q\left(\Ebf + \frac{1}{c}\vbf\wedge\Bbf\right) \ ,
\end{equation}  
\noindent is valid only when the particle charge $q$ is negligible with respect to the
external field sources.  Otherwise, the classical problem of the motion of a
(non-negligible) charge in an electromagnetic field is still an open
question.\cite{quellibroinglese} \
For instance, after the known attempts by Abraham and Lorentz, in 1938
Dirac\cite{DIRA3} obtained and proposed his famous classical equation
\begin{equation}
m \frac{\drm u_{\mu}}{\drm s} = F_{\mu} + {\it\Gamma}_{\mu} \ ,
\label{dirrel}
\end{equation}  
\noi where ${\it\Gamma}_{\mu}$ is the Abraham 4-vector
\begin{equation}
{\it\Gamma}_{\mu}=\frac{2}{3}\frac{e^2}{c}\left({\frac{\drm^2 u_{\mu}}{\drm s^2}
+\frac{u_{\mu}u^{\nu}}{c^2} \frac{\drm ^2u_{\nu}}{\drm s^2}}\right) \ ,
\end{equation}  
\noi that is, the (Abraham) reaction force acting on the electron itself;  and
$F_{\mu}$ is the 4-vector which represents the external field acting on
the particle:
\begin{equation}
F_{\mu}=\frac{e}{c} F_{\mu \nu} u^{\nu} \ .
\end{equation}  
\noindent At the non-relativistic limit, Dirac's equation goes formally into
the one previously obtained by Abraham--Lorentz:
\begin{equation}
m_0\frac{\drm\vbf}{\drm t}-\frac{2}{3}\frac{e^2}{c^3}\frac{\drm^2\vbf}
{\drm t^2}=
e\left(\Ebf +\frac{1}{c}\vbf\wedge\Bbf\right) \ .
\label{dirnrel}
\end{equation}     
\noindent The last equation shows that the reaction force equals \ $\displaystyle {2 \over 3} \; {e^2
\over c^3} \; {\drm^2\vbf \over {\drm} t^2}$.

Dirac's dynamical equation (2) is known to present, however, many troubles,
related with the infinite many non-physical solutions that it possesses. \ Actually, it is
a third--order differential equation, requiring three initial conditions for
singling out a solutions of its. \ In the description of a {\em free} electron,
e.g., it yields ``self-accelerating" solutions ({\em run-away
solutions\/}), for which velocity and acceleration increase
spontaneously and indefinitely. Moreover, for an electron submitted to an
electromagnetic pulse, further non-physical solutions appear, related this
time to {\em pre-accelerations}: If the electron comes from
infinity with a uniform velocity $v_0$ and, at a certain instant of time
$t_0$, it will be submitted to an electromagnetic pulse, then it starts accelerating
{\em before} $t_0$. \ Drawbacks like these motivated further attempts to find
out a coherent (not pointlike) model for the classical electron.

\h Considering elementary particles as points is probably the sin plaguing
modern physics (a plague that, unsolved in classical physics, was transferred
to quantum physics).  One of the simplest way for associating a discreteness
with elementary particles (let us consider, initially, an electron) is just by
the introduction (not of a ``time-lattice", but merely, following
ref.\cite{CALREV4}) of a chronon. We are going to see that, like Dirac's,
Caldirola's approach is also Lorentz invariant (``continuity", in
fact, is not an assumption required by Lorentz invariance). \ Let us
postulate the existence of a universal interval $\tau_0$ of {\em proper}
time, even if time flows continuously as in the ordinary theories.  When an
external force acts on the electron, however, the reaction of the particle
to the applied force is not continuous: The value of the electron velocity
$u_\mu$ is supposed to jump from $u_\mu(\tau - \tau_0)$ to $u_\mu(\tau)$
{\em only at certain positions} $s_{n}$ along its world line; {\em these
``discrete positions" being such that the electron takes a time $\tau_0$ for
travelling from one position $s_{\rm{n} - 1}$ to the next one $s_{n}$.} \ The electron,
in principle, is still considered as pointlike, but the Dirac relativistic
equations for the classical radiating electron are replaced: \ (i) by a
corresponding {\em finite--difference} (retarded) equation in the velocity
$u^\mu(\tau)$
\begin{eqnarray}
{{m_0} \over {\tau_0}}\left\{ {u_\mu \left( \tau  \right)-u_\mu \left(
{\tau -\tau_0} \right)+{{u_\mu \left( \tau  \right)
u_\nu \left( \tau  \right)} \over {c^2}}\left[ {u_\nu \left( \tau
\right)-u_\nu \left( {\tau -\tau_0} \right)} \right]} \right\} = 
{e \over c}F_{\mu \nu}\left( \tau  \right)u_\nu \left( \tau  \right)\,,
\end{eqnarray}   
\noindent which reduces to the Dirac equation (2) when $\tau_{0}$ is small
w.r.t. $\tau$; \ and \
(ii) by a second equation [the {\em transmission law\/}], connecting this time
the discrete positions $x^\mu(\tau)$ along the world line of the particle
among themselves:\\

\hfill{$
x_\mu \left( {n\tau_0} \right)-x_\mu \left[ {\left( {n-1} \right)\tau_0} \right]=
{\displaystyle{\tau_0\over 2}}\left\{ {u_\mu \left( {n\tau_0} \right)-u_\mu \left[ {\left( {n-1}
\right)\tau_0} \right]} \right\} ,
$\hfill}  (6') \\   

\noindent which is valid inside each discrete interval $\tau_{0}$ and describes the
{\em internal} motion of the electron. \ In such equations, $u^\mu(\tau)$ is
the ordinary 4-vector velocity, satisfying the condition \ $u_\mu(\tau)
u^\mu(\tau) = -c^2$ \ for \ $\tau = n \tau_0$, \ where $n = 0,1,2,...$ \ and \
$\mu,\nu = 0,1,2,3$; \ while $F^{\mu \nu}$ is the external (retarded)
electromagnetic field tensor, \ while the chronon associated with the electron
(by comparison with Dirac's equation) results to be, {\it in the simple case} of
an electron interacting with an external field [with $k \equiv (4\pi
\varepsilon_0)^{-1}$],\\

\hfill{$\dis
{\tau_0 \over 2} \equiv \theta_0 = {2 \over 3}{{k e^2} \over {m_0 c^3}} \simeq
6.266 \times 10^{-24} \; {\rm s} \ ,
$\hfill} \\

\noindent depending, therefore, on the particle (internal) properties [namely, on its
charge $e$ and rest mass $m_0$]. \ Things would become different when
considering, e.g., an electron interacting with a macroscopic object,
like in the measurement processes.\cite{DecohRecami}

\h As a result, the electron happens to appear eventually as an
extended--like\cite{RECSAL} particle, with internal structure, rather than
as a pointlike object.  For instance, one may imagine that the particle
does not react instantaneously to the action of an external force because
of its finite extension (the numerical value of the chronon, in the above case,
is just of order of the time spent by light to travel along an electron classical
diameter). \ As already mentioned, eq.(6) describes the motion of an object that
happens to be pointlike only at discrete positions $s_{n}$ along its
trajectory; even if both position and velocity are still continuous and
well-behaved functions of the parameter $\tau$, since they are differentiable
functions of $\tau$.  It is essential to notice that a discrete character
is given in this way to the electron without any need of a ``model" for the
electron.  Actually it is well-known that many difficulties are met not only
by the strictly pointlike models, but also by the extended-{\em type} particle
models (``spheres", ``tops", ``gyroscopes", etc.).  We deem the answer to
stay with a third type of models, the extended-{\em like} ones, as the present
approach; or as the (related) theories in which the center of the {pointlike}
charge is spatially distinct from the particle center-of-mass.\cite{RECSAL} \
Let us repeat, anyway, that also the worst troubles met
in quantum field theory, like the presence of divergencies, are probably
due to the pointlike character still attributed to (spinning) particles; since,
as we already remarked, the problem of a suitable model for elementary
particles was brought, unsolved, from classical to quantum physics.
One might consider that problem to be one of the most important even in modern particle
physics.

\h Equations (6) and the following one provide, together, a full description of
the motion of the electron; and they result to be {\em free} from pre-accelerations,
self-accelerating solutions, and problems with the hyperbolic motion.

In the {\em non-relativistic limit} the previous (retarded) equations
get simplified, into the form

\begin{equation}
{{m_0} \over {\tau_0}}\left[ {\vbf\left( t \right)-\vbf\left(
{t-\tau_0} \right)} \right]= e \left[\Ebf\left( t \right)+{1 \over c}
\vbf\left( t \right)\wedge\Bbf\left( t \right)\right] ,
\end{equation} \\  

\hfill{$
\rbf\left( t \right)-\rbf\left( {t-\tau_0} \right)={\displaystyle{\tau_0\over 2}}\left[
{\vbf\left( t \right)-\vbf\left( {t-\tau_0} \right)} \right] \ .
$\hfill} (7')  \\  

\noindent The important point is that eqs.(6), \ or eqs.(7), \ bypass the
difficulties met by the Dirac classical equation. \ In fact, the
electron {\em macroscopic} motion gets now completely determined, once velocity and
initial position, only, are given. \ The explicit solutions to the above
relativistic equations for the radiating electron  ---or to the corresponding
non-relativistic equations--- solve, indeed, the following questions:

A) {\em case of exact relativistic solutions\/}: \ \  1) free
electron motion; \ 2) electron under the action of an electromagnetic
pulse; \ 3) hyperbolic motion;

B) {\em case of non-relativistic approximate solutions\/}: \ \ 4) electron
under the action of time-dependent forces; \ 5) electron in a constant,
uniform magnetic field; \ 6) electron moving along a straight line under
the action of an elastic restoring force.

\h An explicit study of the electron radiation properties, as deduced
from the finite-difference relativistic equations (6), and their series
expansions, has been carried out by us in refs.\cite{RuyRecami},
showing in detail the advantages of the present formalism
w.r.t. the Abraham-Lorentz-Dirac one.

\subsection{Three alternative formulations}

Two more (alternative) formulations are possible of Caldirola's equations,
based on different discretization procedures. In fact, equations
(6) and (7) describe an intrinsically radiating
particle.  And, by  expanding equation (6))
in terms of $\tau_0$, a radiation reaction term appears:  Those equations
may be called the {\em retarded} form of the electron equations of the motion.

On the contrary, by rewriting the finite--difference equations in
the form

\begin{eqnarray}
{{m_0} \over {\tau_0}}\left\{ {u_\mu \left( {\tau +\tau_0} \right)-u_\mu
\left( \tau \right)+{{u_\mu \left( \tau  \right)u_\nu \left( \tau  \right)}
\over {c^2}}\left[ {u_\nu \left( {\tau +\tau_0} \right)-u_\nu \left( \tau
\right)} \right]} \right\} = {e \over c}F_{\mu\nu}
\left(\tau \right)u_\nu \left(\tau\right)\,,
\end{eqnarray} \\  

\hfill{$
x_\mu \left[ {\left( {n+1} \right)\tau_0} \right]-x_\mu \left( {n\tau_0}
\right)=\tau_0 u_\mu \left( {n\tau_0} \right)\,,
$\hfill} (8')  \\  

\noindent one gets the {\em advanced} formulation of
the electron theory, since the motion is now determined by advanced
actions.  At variance with the retarded formulation, the advanced one
describes an electron which absorbs energy from the external world.

Finally, by adding together retarded and advanced actions, one can
write down the {\em symmetric} formulation of the electron theory:

\begin{eqnarray}
{{m_0} \over {2\tau_0}}\left\{ {u_\mu \left( {\tau +\tau_0} \right)-u_\mu \left(
{\tau -\tau_0} \right)+{{u_\mu \left( \tau  \right)u_\nu \left( \tau  \right)} \over {c^2}}
\left[ {u_\nu \left( {\tau +\tau_0} \right)-u_\nu \left( {\tau -\tau_0} \right)} \right]}
\right\} = {e \over c}F_{\mu\nu}(\tau)u_\nu(\tau)\,, \ \
\end{eqnarray} \\  

\hfill{$
x_\mu \left[ {\left( {n+1} \right)\tau_0} \right]-x_\mu \left( {\left( {n-1} \right)\tau_0}
 \right)=2\tau_0u_\mu \left( {n\tau_0} \right) \ ,
$\hfill} (9')   \\ 

\noindent which does not include any radiation reactions, and describes
a non-radiating electron.\\

\h Before ending our introduction to the classical chronon theory,
let us mention at least one more result derivable from it. \ If one considers a free
particle and look for the ``internal solutions" of  equation (7'),
one gets  ---for a periodical solution of the type
$$
\dot{x}=-\beta_0 \; c \; \sin\left({\frac{2 \pi \tau}{\tau_0}}\right); \ \ \
\dot{y}=-\beta_0 \; c \; \cos\left({\frac{2 \pi \tau}{\tau_0}}\right); \ \ \
\dot{z}=0
$$
\noindent (which describes a uniform circular motion) and by imposing the kinetic energy
of the internal rotational motion to equal the intrinsic energy $m_0c^2$ of
the particle---  that the amplitude of the oscillations is given by
$\beta_0^2=\displaystyle\frac{3}{4}$. \ Thus, the magnetic moment corresponding to this
motion is exactly the {\em anomalous magnetic moment} of the
electron, obtained in a purely classical context: \ $\displaystyle\mu_a=\frac{1}{4 \pi} \;
\frac{e^3}{m_0c^2}$. This shows, by the way, that the anomalous magnetic moment
is an intrinsically classical, and not quantum, result; and the absence of
$\hbar$ in the last expression seems a confirmation of this fact.

\h As to the three {\em interesting} formulations that can be analogously
constructed in the quantum version of the chronon approach, let us here
confine ourselves at quoting refs.\cite{RuyRecami,DecohRecami}.

\section{Spin in classical mechanics}

In some recent papers of ours\cite{RECSAL}, it was proposed a classical
symplectic theory for extended-like\footnote{As already mentioned, the term
{\em extended-like} refers in our language to spinning systems which, even if not
``materially'' extended, nevertheless are something {\em half-way}
between a point and a rotating body (as, e.g., a top). In fact, let us
repeat, in our approaches the center of mass and the
center of charge are distinct points, so that velocity and momentum are not
parallel vectors, and one meets an internal microscopic motion (the so-called
Zitterbewegung\cite{ZBW,Schroedinger}) besides the external one.}
particles, accounting for spin and Zitterbewegung.
\ In particular, in ref.\cite{NNM}, the classical motion of spinning
particles has been described without recourse to any particular models or
formalisms (for instance, without any need of Grassmann variables, Clifford
algebras, or classical spinors), but simply by generalizing the standard
spinless theory. It was only assumed invariance with respect to the Poincar\'e
group, and, from the conservation of the linear and angular momenta, we
derived the Zitterbewegung and the other kinematical properties and motion
constraints. One of us, in ref.\cite{NNM}, has called Non-Newtonian
Mechanics (NNM) such a classical approach. Indeed, newtonian mechanics is
re-obtained as a particular case: namely, for spinless systems with no
Zitterbewegung.

\h Let us start from a Poincar\'e-invariant Lagrangian, which generalizes the
newtonian Lagrangian $\Lczero=\um mv^2$ (where $v^2\equiv v_\mu
v^ \mu$) by means of proper-time derivatives of the velocity up to the $N$-th
order (when the scalar potential is $U=0$, and only free particles are considered):
\bb    
\LcN \equiv \um Mv^2 + \um k_1\dov^2 + \um k_2\ddov^2 + \cdots -
\equiv \sum_{n=0}^N\,\um k_n^N{v^{(\rm n)}}^2\,,
\label{Lagr(N)}
\ee
where the notation $^{(n)}$ indicates the $n-$th derivative with respect to
$\tau$ and the $k_n^N$ are scalar coefficients (endowed with alternate signs).
The Euler-Lagrange equation of the motion
\bb
\frac{\pa\Lc}{\pa x} = \frac{\rd}{\dtau}\left({\frac{\pa\Lc}{\pa\dox}}\right) -
\frac{\rd^2}{\dtau^2}\left({\frac{\pa\Lc}{\pa\ddox}}\right) +
\frac{\rd^3}{\dtau^3}\left({\frac{\pa\Lc}{\pa\dddox}}\right) -
\cdots
\ee
yields a constant-coefficient $N$-th order differential equation, which
can be regarded as a generalization of Newton law ($F^\mu = Ma^\mu$), in
which non-newtonian {\em Zitterbewegung terms} appear:
\bb
0 \ug Ma^\mu + \sum_{n=1}^N\,(-1)^{{\rm n}}k_n^N\,\aii^\mu\,, \label{GNEq}
\ee
where we have put here $F=0$ since we assumed $U=0$. Incidentally, alternate
signs for the coefficients of the terms appearing in the Lagrangian are
requested if we want stationary solutions and finite oscillatory
motions only.

\h The zero-th order canonical momentum \
$$\dis\frac{\pa\Lc}{\pa\dox_\mu} -
\frac{\rd}{\dtau}\left({\frac{\pa\Lc}{\pa\ddox_\mu}}\right) +
\frac{\rd^2}{\dtau^2}\left({\frac{\pa\Lc}{\pa\dddox_\mu}}\right) -
\cdots \,,
$$
conjugate to $x^\mu$, writes:
\bb
p_{[0]}^\mu =  Mv^\mu + \sum_{n=1}^N\,(-1)^{{\rm n}}\,k_n^N\,{v^{(\rm 2n)}}^\mu =
\sum_{n=0}^N\,(-1)^{{\rm n}}\,k_n^N\,{v^{(\rm 2n)}}^\mu\,.
\label{ZeroMom}
\ee
By imposing the symmetry of the Lagrangian under 4-rotations, one gets
the conservation of the total angular momentum, which results to be
composed of the usual orbital angular momentum tensor and of a classical
spin tensor, defined by employing classical kinematical quantities only.
Indeed, via the Noether theorem, the spin tensor and the spin vector can
be written as follows:
\bb
S_\munu \ug \sum_{n=1}^Nk_n^N\,\sum_{l=0}^{n-1}(-1)^{n-l-1}
\left(v_\mu^{(l)}v_\nu^{(2n-l-1)}-v_\nu^{(l)}v_\mu^{(2n-l-1)}\right)\,;
\ee
\bb
\sbf \ug \sum_{n=1}^Nk_n^N\,\sum_{l=0}^{n-1}(-1)^{n-l-1}
\vbf^{(l)}\times\vbf^{(2n-l-1)}\,,           \label{NNMspin}
\ee
respectively.

\h As an example let us take $N=4$.  One has
\bb
\imp \ug M\vbf-k_1\doabf+k_2\dddoabf-k_3\abf^{({\rm V})}+
k_4\abf^{({\rm VII})}\,.
\ee
The spin is
$$
\sbf \ug k_1(\vbf\times\abf) + k_2(\abf\times\doabf-\vbf\times\ddoabf)
+k_3\left(\doabf\times\ddoabf-\abf\times\dddoabf+
\vbf\times\abf^{({\rm IV})}\right)+
$$
$$
+k_4\left(\ddoabf\times\dddoabf-\doabf\times\abf^{({\rm IV})}+
\abf\times\abf^{({\rm V})}- \vbf\times\abf^{({\rm VI})}\right)\,;
$$
thus, after differentiating and simplifying, we obtain
\bb
\dosbf \ug \vbf\times\left(k_1\doabf-k_2\dddoabf+k_3\abf^{({\rm V})}-
k_4\abf^{({\rm VII})}\right) = \vbf\times(M\vbf-\imp)=\imp\times\vbf\,,
\ee
as expected.\footnote{From the conservation of the total angular momentum
$\jbf$, the sum of the orbital and spin angular momentum, that is, from
relation
$$
\dot{\jbf}=\dot{\lbf}+\dosbf=0 \,,      
$$
we actually get, by taking into account also the momentum conservation, 
$\dot{\imp}=0$, that
$$
\dis\dosbf=-\dot{\lbf}=-\frac{\rd}{\dtau}(\xbf\times\imp)=-\vbf\times\imp\,. 
$$}

\h The hamiltonian representation of the theory is obtained by
introducing, besides the (constant) zero-th order momentum $p_{[0]}^\mu$
given by eq.(\ref{ZeroMom}), the other (non-constant) $l$-th order momenta
$p_{[l]}^\mu$ canonically conjugate to $x_{[l]}\equiv x^{(l)}$
\bb
p_{[l]}^\mu \equiv \sum_{n=l}^N(-1)^{n-l}\frac{\rd^{n-l}}{\dtau^{n-l}}
\left(\frac{\pa\Lc}{\pa x^{(n+1)}}\right)
= \sum_{n=l}^N(-1)^{n-l}k_n^Nv^{(2n-l)}    \label{l-momentum}
\ee
[a definition which includes also the $l=0$ case, Eq.\,(\ref{ZeroMom})].
On employing the high order momenta, the spin vector (\ref{NNMspin})
can be put in the canonical form
\bb
\sbf \ug \sum_{l=1}^N\xbf_{[l]}\times\imp_{[l]}\,, \label{spincanonic}
\ee
analogous to that of the orbital angular momentum, $\lbf=\xbf\times\imp_{[0]}$.

\h The {\em conserved} scalar Hamiltonian, obtained by imposing the
$\tau$-reparametrization invariance of the Lagrangian, is
\bb
\Hc = \sum_{l=0}^Np_{[l]}^\mu\dox_{[l]\mu} - \Lc =
\um Mv^2 + \sum_{n=1}^Nk_n^N\,\left[\um{v^{(n)}}^2 +
\sum_{l=0}^{n-1}(-1)^{n-l}v^{(l)\mu}v^{(2n-l)}_\mu\right]\,.
\label{NNMscalar}
\ee
\h It can be also shown that a couple of Hamilton equations
\bb
\dox_{[l]}^\mu \ug \frac{\pa\Hc}{\pa p_{[l]\mu}}  \qquad\qquad\quad
\dop_{[l]}^\mu \ug -\frac{\pa\Hc}{\pa x_{[l]\mu}}
\ee
holds for any couple of canonical variables $\dis\left(x_{[l]}^\mu;
\ p_{[l]}^\mu\right)$, and that the set of the Hamilton equations is
globally equivalent to the Euler-Lagrange equation (\ref{GNEq}).
\h The Poisson brackets are here defined as follows
\bb
\dis\{f,g\} \equiv \sum_{l=0}^N\left(\frac{\pa f}{\pa x_{[l]}^\mu}
\frac{\pa g}{\pa p_{[l]\mu}} -
\frac{\pa f}{\pa p_{[l]}^\mu}\frac{\pa g}{\pa x_{[l]\mu}}\right)\,;
\ee
while the time evolution of a generic quantity is given by
\bb
\dot{G} \ug \frac{\pa G}{\pa t} + \{\Hc, G\}\,.
\ee
\h The action $\Sc=\dis\int\Lc\dtau$ can be written in the characteristic form
\bb
\Sc \ug \sum_{l=0}^N\int p_{[l]}^\mu\rd x_{[l]\mu} - \int\Hc\dtau\,,
\ee
from which one has:
\bb
p_{[l]}^\mu=\frac{\pa\Sc}{\pa x_{[l]\mu}} \qquad\qquad
\Hc=-\,\frac{\pa\Sc}{\pa\tau}\,.
\ee

\

\section{Spin and Hamiltonian in chronon theory}

In the symmetric formulation of the chronon theory,
Caldirola's Lagrangian\cite{Caldirola} writes (in the case of free
particles) as
\bb
\Lc \ug
\sum_{n=0}^\infty\um\,\frac{(-1)^nM\tz^{2n}}{(2n+1)!}\,{v^{(\rm n)}}^2\,.
\label{Lagrsimm}
\ee
It does coincide, therefore, with the infinite-order ($N\to\infty$)
non-newtonian Lagrangian (\ref{Lagr(N)}), provided that one assumes
$$
k_n\equiv\frac{(-1)^nM\tz^{2n}}{(2n+1)!}\,.
$$
We can say that such a {\em symmetric} chronon theory can be regarded just as a
particular case of NNM, entailing a periodic motion, endowed a priori with
all the infinite harmonics of the ground frequency $\omz=2\pi/2\tz=\pi/\tz$.

\h We shall prove that, in the absence of external fields, the (total) velocity $v^\mu$ can be expressed by
a generic periodic function expanded in Fourier series (in the following,
$E_m^\mu$ and $H_m^\mu$ are arbitrary
constant spacelike 4-vectors fixing the ``internal'' initial conditions, whilst
$p^\mu$ fixes the ``external'' one\cite{RECSAL,Schroedinger,NNM,Spincord})\footnote{Solution (\ref{velocity}) with \ $E_m, H_m \neq 0$ \ holds only for \ $\tz\neq 0$; \ whilst in the ``newtonian'' limit, \ $\tz\to 0$, \ the (free particle) motion equations reduce, of course, to \ $a^\mu=0, v^\mu=p^\mu/M$, see Eq.(\ref{EL-eq}).}
\bb\
v^\mu \ug \frac{p^\mu}{M} + \sum_{m=1}^\infty E_m^\mu\cos(m\omz\tau)+
H_m^\mu\sin(m\omz\tau)\,.         \label{velocity}
\ee
\h Let us recall that the most general representations of the Lorentz group
imply that the space-time rotation operator $J^\munu$ is (already classically)
decomposed into an orbital part and a spin part, $J^\munu=L^\munu +
S^\munu$. Incidentally, the phase-space doubles, and the 4-rotations
parameters are $6+6=12$. As a consequence, the motion consists in two
parts, the translational (external, newtonian) and the rotational
(internal) term; and correspondingly the total velocity $v$ consists in
the two parts $v=w+V$. The former, i.e. the drift velocity $w \equiv p^\mu/M$,
corresponds to the center-of-mass motion, namely to the average motion
of the considered particle; by definition, since $p^2=M^2$, one gets that
as usual $w^2=1$. One meets only subluminal speeds for such an external
speed, which expresses the 4-momentum propagation speed, and is the only one
till now experimentally observed. The latter,
$V= \sum_{m=1}^\infty E_m^\mu\cos(m\omz\tau)+H_m^\mu\sin(m\omz\tau)$,                                                       
is directly linked to the discrete time approach, and simultaneously happens
to describe the spin motion.

Notice that, on condition that the time parameter be just the time referred to the 
center-of-mass frame, the total 4-velocity squared $v^2$ turns out to be
{\em not} constrained to 1, not only in the present theory, but also in many other recent 
works describing spinning ``rigid'' particles\cite{Pavsic,Plyushchay,Polyakov}
where the the classical action contains additional terms dependent on the so-called 
``extrinsic curvature'' (that is, on the 4-acceleration squared).\footnote{Classical 
equations of the motion for a rigid particle or for a rigid $n-$dimensional worldsheet, 
either in flat or curved background spacetimes, have been derived from Lagrangians containing
also terms dependent on higher derivatives of the 4-velocity (``torsion''-terms,
etc.). In \cite{Nesterenko} the equations of motion are reformulated in terms of the 
principal wordline curvatures which result to be motion integrals, namely mass and
spin. As it occurs in NNM, in the mentioned approaches the total velocity squared is in general a function of mass and spin as well.} \ \h Let us also recall that the kinematical explicit derivation of the spin vector
from a classical Lagrangian, presented in this paper for the first time
within a discrete-time theory, is a very recent result obtained by us and very few
others\cite{RECSAL,NNM,Pavsic,Plyushchay,Polyakov}. The previous authors engaged 
with the chronon approaches (see, e.g., ref.\cite{Caldirola}) did not notice the 
natural emergence of spin from time discreteness.

\h The solution (\ref{velocity}) satisfies the Euler-Lagrange equation for
Lagrangian (\ref{Lagrsimm}), that is to say
\bb
Ma+M\frac{\tz^2}{3!}\ddoa+M\frac{\tz^4}{5!}a^{({\rm IV})}+\ldots=
\sum_{n=0}^\infty\frac{M\tz^{2n}}{(2n+1)!}\,a^{(2n)}=0 \,,
\label{EL-eq}
\ee
as well as the (relativistic and nonrelativistic) Caldirola equations in the
non-symmetrical (advanced or retarded) formulations, in the absence
of electromagnetic fields. This can be easily proved as follows.
Solution (\ref{velocity}) is periodic with period $\dis 2\tz$;
then we have for any $\tau$
$$
v_\mu(\tau+\tz)-v_\mu(\tau-\tz)\ug 0\,.
$$
On expanding in Taylor series, one gets
$$
\sum_{n=0}^\infty\frac{v_\mu^{(n)}(\tau)}{n!}\tz^n \,-\,
\sum_{n=0}^\infty(-1)^n\frac{v_\mu^{(n)}(\tau)}{n!}\tz^n \ug 0
$$
and then
\bb
\sum_{n=0}^\infty\frac{v_\mu^{(2n+1)}(\tau)}{(2n+1)!}\tz^{2n+1} \equiv
\sum_{n=0}^\infty\frac{a_\mu^{(2n)}(\tau)}{(2n+1)!}\tz^{2n+1} = 0\,,
\ee
from which eq.(\ref{EL-eq}) follows.

\

\h Let us find the explicit expression of the spin vector in the
chronon theory, choosing for convenience the center-of-mass (where $\imp=0$)
as the reference frame. By inserting equation (\ref{velocity}) with $\imp=0$ 
into equation
$$
\sbf \ug \sum_{n=1}^\infty k_n\,\sum_{l=0}^{n-1}(-1)^{n-l-1}
\vbf^{(l)}\times\vbf^{(2n-l-1)}
$$
[which is nothing but eq.(\ref{NNMspin}) for $N\to\infty$],
after some algebra one obtains
\bb
\sbf \ug \sum_{m=1}^\infty A_m\,(\Ebf_m\times\Hbf_m)
\ee
with
\bb
A_m \equiv \um\sum_{n=1}^\infty n\kb_nm^{2n}\,,
\ee
where the dimensionless coefficients $\kb_n$ are defined as follows:
\bb
\kb_n \equiv k_n\frac{\omz^{2n}}{M}\,.
\ee
\h By exploiting the explicit expression of $\kb_n$, we have
$$
\sum_{n=0}^\infty \kb_n x^{2n} =
\frac{1}{\pi x}\sum_{n=0}^\infty\frac{(-1)^n\pi^{2n+1}}{(2n+1)!}x^{2n+1}
= \frac{\sin(\pi x)}{\pi x}\,.
$$
Differentiating side by side the above equation,
we obtain
$$
\sum_{n=0}^\infty n\kb_n x^{2n-1} =
\frac{\cos(\pi x)}{2x}-\frac{\sin(\pi x)}{2\pi x^2}\,,
$$
which, after multiplication of both its sides by $x$, yields
for $x=m\in\mathbb{N}^+\;$:
\bb
\sum_{n=0}^\infty n\kb_n m^{2n} = \frac{\cos(\pi m)}{2} =
\frac{(-1)^m}{2}\,.     \label{sumpesata}
\ee
The property
\bb
\sum_{n=0}^\infty\kb_n m^{2n} \ug 0 \label{sumkbm}
\ee
holds also for any positive integer $m$. \
Indeed, this relation can be obtained by putting $x$ equal to
$m\in\mathbb{N}^+$ into the equation
$$
\sum_{n=0}^\infty \kb_n x^{2n} = \frac{\sin(\pi x)}{\pi x}\,.
$$
\h Taking into account Eq.\,(\ref{sumpesata}), the spin vector can be
eventually written as follows:
\bb
\fbox{${\dis\sbf \ug
\frac{1}{4}\sum_{m=1}^\infty(-1)^m\Ebf_m\times\Hbf_m}$}\,.
\ee
We can proceed quite analogously in order to get the Hamiltonian
in the chronon theory, in a generic reference-frame. By exploiting
eq.(\ref{NNMscalar}), with $N\to\infty$,
$$
\Hc = \um Mv^2 + \sum_{n=1}^\infty k_n\,\left[\um{v^{(n)}}^2 +
\sum_{l=0}^{n-1}(-1)^{n-l}v^{(l)\mu}v^{(2n-l)}_\mu\right]\,,
$$
after some calculations we get \ ($E^2\equiv E_\mu E^\mu$;
$H^2\equiv H_\mu H^\mu$) \ that
\bb
\Hc \ug \frac{p^2}{2M} + M^3\sum_{m=1}^\infty B_m(E_m^2+H_m^2)\,,
\ee
with
\bb
B_m \equiv 1+2\sum_{n=1}^\infty\left(n+\um\right)\kb_nm^{2n}\,.
\ee
\h Taking into account eqs.(\ref{sumpesata}) and (\ref{sumkbm}), we
finally have:
\bb
\fbox{${\dis\Hc\ug \frac{p^2}{2M} +
M^3\sum_{m=1}^\infty[1+(-1)^m](E_m^2+H_m^2)}$}\,,
\ee
where, besides the ordinary ``external" drift term $\dis\frac{p^2}{2M}$, it
appears an ``internal" term, which seem to constitute a signature of the
actual ``non-newtonian nature" of the chronon theory.

\section{Conclusions}

In this paper we have analytically derived the spin vector and the Hamiltonian
for a {\em non-quantum} theory employing a discretized time: the chronon theory.
We have shown that such an approach is nothing but a particular case of
NNM (``non-newtonian mechanics"), a classical theory where the motion of
the spinning, extended-like particles is described in terms of
an infinite set of time derivatives: The spin arises just from this 
underlying non-local structure.  In spite of the classical character of the
chronon theory, we have obtained an explicit
kinematical formulation of the intrinsic angular momentum (usually considered
a pure {\em quantum} quantity), through a sum over all harmonic modes of the
solution to the motion equation.

\h The present approach to particle dynamics appears as an alternative to the
better known string model, for taking account of the internal degrees of
freedom which generate the rich variety of the observed particles. As a
matter of fact, also the general solution to the string motion equation
consists of a sum over all harmonic modes; further, in ref.\cite{Spincord},
one of us has shown that even bosonic strings obey the
motion equation of the chronon theory, Eq.\,(\ref{EL-eq}). In a sense,
the chronon theory (in particular, Caldirola's) might be considered as having
anticipated, fifty years ago already, the string concept of our days.
In a forthcoming paper we shall try to study more deeply the quantized
version\cite{RuyRecami} Caldirola's theory, in order to obtain 
the allowed spin states and a mass spectrum.\footnote{On this respect,
let us anticipate that, however, in a (second-quantization) quantum field theory,
the quantization follows from the replacement
of the Poisson brackets with the corresponding commutators: So that no
explicit time-discretization will be really needed (as, instead, it occurs
in the first-quantization of the chronon theory, which has to make recourse
to finite-difference Schr\"odinger-like wave equations\cite{RuyRecami}.)}

\

\

\noindent {\bf Acknowledgements} \\
Many thanks are due to Ruy H.A. Farias, for his very useful collaboration
along many years, and to Rodolfo Bonifacio, Michel Zamboni-Rached
for some stimulating discussions or hints. This work
has been partially supported by I.N.F.N. and M.I.U.R.

\


\begin{thebibliography}{99}

\bibitem{GUTs} D.Bailey and A.Love, {\em Supersymmetric Gauge Field Theory and
String Theory}, chapter 6, and references therein, (I.O.P. Publishing;
Bristol, 1994)

\bibitem{String} V.A.Kosteleck\'y, S.Samuel, Phys.Rev. {D39} (1989) 683;
Phys.Rev.Lett. {\bf 63} (1989) 224; {\bf 66} (1991) 1811;
\ V.A.Kosteleck\'y and R.Potting, Nucl.Phys. {\bf B359} (1991) 545;
Phys.Lett. {\bf B381} (1996) 89; Phys.Rev. {\bf D63} (2001) 046007;
\ V.A.Kosteleck\'y, M.J.Perry and R.Potting, Phys.Rev.Lett. {\bf 84}
(2000) 4541; \ C.P.Burgess, JHEP  {\bf 0409} (2004) 033;
JHEP {\bf 0203} (2002) 043; \ A.R.Frey, JHEP {\bf 0304} (2003) 012;
\ J.Cline and L.Valc\'arcel, e-print ph/0312245;
\ F.Lizzi and R.J.Szabo, Comm.Math.Phys. {\bf 197} (1998) 667;
\ F.Lizzi, G.Mangano, G.Miele, JHEP {\bf 06} (2002) 049;
\ M.Kalb and D.Ramond, Phys.Rev. {\bf D9} (1974) 2273;
\ E.Cremmer and J.Scherk, Nucl.Phys. {\bf 72} (1974) 117;
\ Y.Nambu, Phys.Rep. {\bf C23} (1976) 251

\bibitem{QG} F.R.Klinkhamer, Nucl.Phys. {\bf B578} (2000) 277; \ J.Alfaro,
H.A.Morales-T\'ecotl and L.F.Urrutia, Phys.Rev. {\bf D66} (2002) 124006;
\ D.Sudarsky, L.Urrutia, and H.Vucetich, Phys.Rev.Lett. {\bf 89} (2002) 231301;
Phys.Rev. {\bf D68} (2003) 024010; \ F.R.Klinkhamer and C.Rupp, e-print
th/0312032; \ C.J.Isham, arXiv:gr-qc/9510063; \ J.Butterfield and C.J.Isham,
e-print gr-qc/9901024;
\ C.Rovelli, e-print gr-qc0006061; {\em Quantum Gravity},  (Cambridge University Press;
Cambridge, 2004); \ G.Amelino-Camelia, Nature {\bf 408} (2000) 661;
C.P.Burgess, Living Rev. Rel. {\bf 7} (2004) 5

\bibitem{Foam} L.J.Garay, Phys.Rev.Lett. {\bf 80} (1998) 2508;
G.Amelino-Camelia, J.R.Ellis, N.E.Mavromatos and D.V.Nanopoulos, Nature {\bf 12}
(1997) 607

\bibitem{NCG} G.Amelino-Camelia, arXiv:hep-th/0211022; Phys.Lett. {\bf B392} (1997) 283; 
Int.J.Mod.Phys. {\bf D11} (2002) 35; Phys.Lett. {\bf B510} (2001) 255;
\ G.Amelino-Camelia and T.Piran, Phys.Rev. {\bf D64} (2001) 036005;
\ G.Amelino-Camelia, L.Doplicher, S.Nam and Y.Seo, Phys.Rev. {\bf D67} (2003) 085008;
\ N.R.Bruno, G.Amelino-Camelia and J.Kowalski-Glikman, Phys.Lett. {\bf B522} (2001) 133;
\ S.X.Chen and Z.Y.Yang, Mod.Phys.Lett. {\bf A18} (2003) 2913;
\ Z.Y.Yang and S.X.Chen, J.Phys. {\bf A35} (2002) 9731;
\ J.Lukierski, H.Ruegg and W.J.Zakrzewski, Ann.Phys. {\bf 243} (1995) 90;
\ Z.Guralnik, R.Jackiw, S.Y.Pi and A.P.Polychronakos, Phys.Lett. {\bf B517} (2001) 450;
 \ V.Nazaryan  and C.E.Carlson, Phys.Rev. {\bf D71} (2005) 025019;
\ C.E.Carlson, C.D.Carone and R.F.Lebed, Phys.Lett. {\bf B549} (2002) 337;
\ M.Hayakawa, Phys.Lett. {\bf B478} (2000) 394; arXiv:hep-th/9912167; \ S.M.Carroll,
J.A.Harvey, V.A.Kosteleck\'y, C.D.Lane and T.Okamoto, Phys.Rev.Lett. {\bf 87} (2001) 141601;
\ A.Anisimov, T.Banks, M.Dine and M.Graesser, Phys.Rev. {\bf D65} (2002) 085032;
\ T.Kifune, Astrophys.J.Lett. {\bf L518} (1999) 21;
\ W.Kluzniak, arXiv:astro-ph/9905308; \ R.J.Protheroe and H.Meyer, Phys.Lett.
{\bf B493} (2000) 1; \ D.Bahns, S.Doplicher, K.Fredenhagen and G.Piacitelli,
Phys.Rev. {\bf D71} (2005) 025022; \ C.K.Zachos, Mod.Phys.Lett. {\bf A19} (2004) 1483

\bibitem{LQG} R.Gambini and J.Pullin, Phys.Rev. {\bf D59} (1999) 124021;
Phys.Rev. {\bf D65} (2002) 103509; \ G.'t Hooft, Class.Quant.Grav. {\bf 13}
(1996) 1023; \ J.Alfaro, H.A.Morales-Tecotl and L.F.Urrutia, Phys.Rev.Lett.
{\bf 84} (2000) 2318; \ C.Rovelli and L.Smolin, Phys.Rev. {\bf D52} (1995) 5743;
Nucl.Phys. {\bf B442} (1995) 593; Erratum-ibid. {\bf B456} (1995) 753

\bibitem{CALREV4} P.Caldirola, Nuovo Cimento {\bf 10} (1953) 1747;
\ {\em Rivista Nuovo Cim.} {\bf 2} (1979), issue no.13, and refs. therein;
{\em Revista Brasil. de F\'{\i}sica}, special volume (1984, July), p.228.
\ See also R.Cirelli,
Nuovo Cimento {\bf 1} (1955) 260; \ L.Lanz, Nuovo Cimento {\bf 23}
(1962) 195; \  F.Casagrande and E.Montaldi, Nuovo Cimento {\bf A40} (1977)
369; {\bf A44} (1978) 453; \ P.Caldirola and E.Montaldi, Nuovo Cimento
{\bf B53} (1979) 291; \ P.Caldirola, G.Casati and A.Prosperetti, Nuovo Cimento
{\bf A43} (1978) 127; \ P.Caldirola, Nuovo Cimento {\bf A49} (1979) 497; \
A.Prosperetti, Nuovo Cimento {\bf B57} (1980) 253; \ L.Belloni, Lett.
Nuovo Cim. {\bf 31} (1981) 131; \ V.Benza and P.Caldirola, Nuovo
Cimento {\bf A62} (1981) 175; \ G.C.Ghirardi and T.Weber: Lett. Nuovo
Cim. {\bf 39} (1984) 157; and in particular R.Bonifacio and P.Caldirola,
Lett. Nuovo Cim. {\bf 38} (1983) 615; {\bf 33} (1982) 197. \ Cf. also
T.D.Lee, ``Can time be a discrete dynamical variable?", Phys. Lett. {\bf B122}
(1983) 217; \ G.Jaroszkiewicz, ``Principles of discrete time mechanics",
J.Phys.A:Math.Gen. {\bf 30} (1997) 3115; {\em ibidem}, 3145; \ {\bf 31} (1998)
977; {\em ibidem}, 1001

\bibitem{quellibroinglese} A.D.Yaghjian: {\em Relativistic Dynamics of a
Charged Sphere} (Springer; Berlin, 1992)

\bibitem{DIRA3} P.A.M.Dirac, ``The classical theory of electron",
Proc. Royal Soc. {\bf A167} (1938) 148; \ Ann. Inst. Poincar\'e
{\bf 9} (1938) 13. \ Cf. also M.Sch\"onberg et al., Phys. Rev. {\bf 69}
(1945) 211; and Anais Ac. Brasil. Cie. {\bf 19} (1947), issue no.3,
pp.46-98

\bibitem{DecohRecami} E.Recami, "A simple quantum equation for dissipation
and decoherence" [report NSF-ITF-02-62 (I.T.P., UCSB; California, 2002)],
in {\em Quantum Computing and Quantum Bits in Mesoscopic Systems}, ed. by
A.J.Leggett, B.Ruggiero and P.Silvestrini (Kluwer/Plenum; New York, 2004),
pp.111-122

\bibitem{RECSAL} G.Salesi, Mod. Phys. Lett. {\bf A11} (1996) 1815;
Int. J. Mod. Phys. {\bf A12} (1997) 5103; \ G.Salesi and E.Recami, Phys. Lett.
{\bf A190} (1994) 137; {\bf A195} (1994) E389; \ Found. Phys. {\bf 28} (1998)
763; \ E.Recami and G.Salesi, Phys. Rev. {\bf A57} (1998) 98; \
Adv. Appl. Cliff. Alg. {\bf 6} (1996) 27; \ in {\em Gravity, Particles
and Space-Time}, ed. by P.Pronin and G.Sardanashvily (World Scient.;
Singapore, 1996), pp.345-368; \ M.Pav\v{s}i\v{c}, E.Recami, W.A.Rodrigues,
G.D.Maccarrone,
F.Raciti and G.Salesi, Phys. Lett. {\bf B318} (1993) 481; \ W.A.Rodrigues,
J.Vaz, E.Recami and G.Salesi, Phys. Lett. {\bf B318} (1993) 623; \
\ J.Vaz and W.A.Rodrigues, Phys. Lett. {\bf B319} (1993) 203

\bibitem{RuyRecami}  R.H.A.Farias and E.Recami, ``Introduction of a quantum of time
(chronon), and its consequences for quantum mechanics", Report IC/98/74
(ICTP; Trieste, 1998), appeared in preliminary form as e-print
quant-ph/9706059. \ Cf. also R.H.A.Farias, ``Introduction of
a `quantum' of time into the formalism of quantum mechanics", PhD Thesis,
E.Recami supervisor (UNICAMP; Campinas, S.P.; 1994)

\bibitem{ZBW} P.A.M.Dirac, {\em The principles of Quantum Mechanics} (Claredon;
Oxford, 1958), $4^{\rm th}$ edition, p.262; \ J.Maddox, Nature  325 (1987) 306

\bibitem{Schroedinger} E.Schr\"{o}dinger, Sitzunger. Preuss. Akad. Wiss.
Phys.-Math. Kl. {\bf 24} (1930) 418; \ {\bf 25} (1931) 1

\bibitem{NNM} G.Salesi, Int. J. Mod. Phys. {\bf A17} (2002) 347
(e-print: quant-ph/0112052); \ {\bf A20} (2005) 2027

\bibitem{Caldirola} P.Caldirola, {\em Suppl. Nuovo Cim.} {\bf 3} (1956) 297

\bibitem{Spincord} G.Salesi, Found.Phys.Lett. {\bf 19} (2006) 367 

\bibitem{Pavsic} M. Pav\v{s}i\v{c}, Phys. Lett. {\bf B205}, 231 (1988); 
{\bf B221}, 264 (1989); Class. Quant. Grav. {\bf L7}, 187 (1990) 

\bibitem{Plyushchay} M.S. Plyushchay, {\em Comment on the relativistic particle with
curvature and torsion of world trajectory}, arXiv:hep-th/9810101; 
Phys. Lett. {\bf  B262}, 71 (1991); Mod. Phys. Lett. {\bf A3}, 1299 (1988);
{\bf A4}, 837 (1989); {\bf A4}, 2747 (1989); Int. J. Mod. Phys. {\bf A4}, 3851 (1989); 
Phys. Lett. {\bf B243}, 383 (1990); Phys. Lett. {\bf B236}, 291 (1990); 
{\bf B235}, 47 (1990); {\bf B253}, 50 (1991)

\bibitem{Polyakov} A.M. Polyakov, Nucl. Phys. {\bf B268}, 406 (1986); Mod. Phys.
Lett. {\bf A3}, 325 (1988);  \ Yu.A. Kuznetsov and A.M. Polyakov, Phys. Lett. 
{\bf B297}, 49 (1992)

\bibitem{Nesterenko} V.V. Nesterenko, A. Feoli, G. Scarpetta, J. Math. Phys. 
{\bf 36}, 5552 (1995); \ V.V. Nesterenko, Phys. Lett. {\bf B327}, 50 (1994)

\end{thebibliography}
\end{document}